\documentclass[conference]{IEEEtran}
\IEEEoverridecommandlockouts
\usepackage{cite}
\usepackage{amsmath,amssymb,amsfonts}
\usepackage{algorithm}
\usepackage{graphicx}
\usepackage{textcomp}
\usepackage{xcolor}
\usepackage{algpseudocode}
\usepackage{multirow}
\usepackage{multicol}
\usepackage{graphicx}
\usepackage{booktabs}
\usepackage{colortbl}

\def\BibTeX{{\rm B\kern-.05em{\sc i\kern-.025em b}\kern-.08em
    T\kern-.1667em\lower.7ex\hbox{E}\kern-.125emX}}
\begin{document}

\title{HPD: \textbf{H}ybrid \textbf{P}rojection \textbf{D}ecomposition for Robust State Space Models on Analog CIM Hardware}
\author{
    \IEEEauthorblockN{Yuannuo Feng\textsuperscript{\dag}$^{1,3}$, Wenyong Zhou\textsuperscript{\dag}$^{2,3}$, Yuexi Lyu\textsuperscript{\ddag}$^{3}$, Hanjie Liu$^{3}$, Zhengwu Liu*$^{2}$, Ngai Wong*$^{2}$, Wang Kang*$^{1}$}
    \IEEEauthorblockA{
        $^1$School of Integrated Circuit Science and Engineering, Beihang University, Beijing, China\\
        $^2$Department of Electrical and Electronic Engineering, The University of Hong Kong, Hong Kong\\
        $^3$Zhicun Research Lab, Beijing, China\\
       \textsuperscript{\dag}: Equal Contribution. \textsuperscript{\ddag}: Project Leader. *: Corresponding Author(s).
    }
}

\maketitle
\begin{abstract}
State Space Models (SSMs) are efficient alternatives to traditional sequence models, excelling at processing long sequences with lower computational complexity. Their reliance on matrix multiplications makes them ideal for compute-in-memory (CIM) architectures, which improve energy efficiency by computing within memory arrays. However, device non-idealities in CIM introduce weight perturbations that can degrade inference accuracy.
In this paper, we systematically analyze the robustness of SSMs under noisy conditions, identifying that the final block and output projection layers are more susceptible to perturbations compared to other components. Building on these insights, we propose HPD, a Hybrid Projection Decomposition strategy for the last output projection layer. We replace the original weight matrix with the multiplication of \( \mathbf{U} \) and \( \mathbf{\Sigma} \) in its SVD to ensure compatibility with existing hardware architectures, while offloading \( \mathbf{V^\top} \) to digital hardware for precise and robust correction.
Comprehensive tests on Mamba models show that our method reduces perplexity by up to 99.57\% under various noise conditions compared to baseline models, with accuracy gains of up to 96.67\% on the PIQA benchmark for commonsense reasoning.
\end{abstract}

\begin{IEEEkeywords}
State Space Models, Compute-in-Memory
\end{IEEEkeywords}
\section{Introduction}
\label{sec:introduction}
\begin{figure}[!t]
\centering
\includegraphics[scale=0.28]{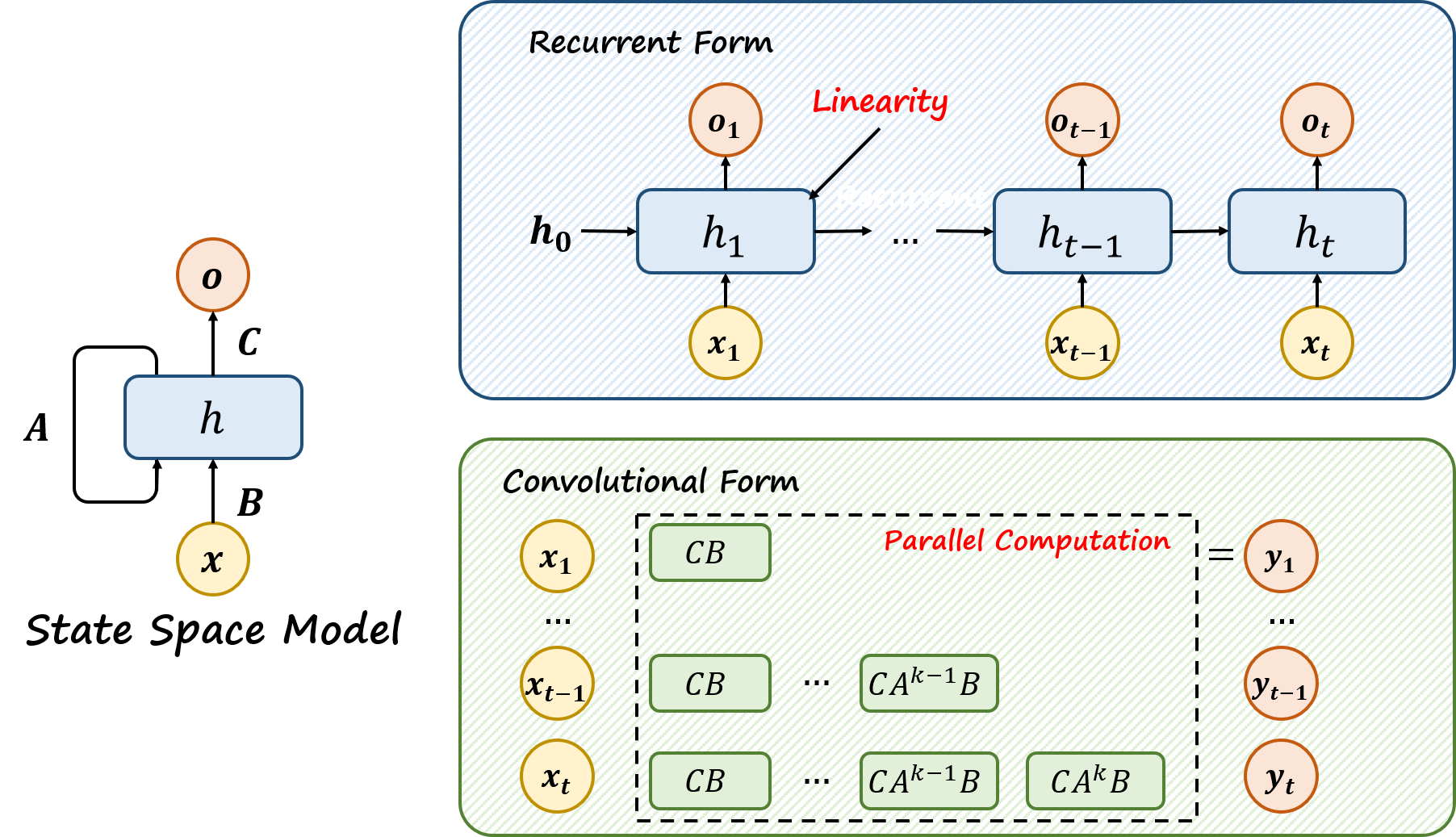}
\caption{An illustration of SSM architectures, combining recurrent and convolutional computations via their linear nature. SSMs enable recurrent inference and parallel training, leveraging RNN and Transformer strengths.}
\label{fig:mamba}
\end{figure}

State Space Models (SSMs) have emerged as efficient alternatives to traditional sequence modeling architectures, with recent variants like Mamba~\cite{gu2023mamba, mamba2} achieving Transformer-competitive performance~\cite{gpt} while using substantially fewer computational resources for long sequences, as shown in Fig.~\ref{fig:mamba}. SSMs achieve their efficiency through structured parameterization that enables linear scaling with sequence length, relying heavily on matrix multiplication operations that are ideally suited for analog compute-in-memory (CIM) acceleration~\cite{icsict, chen2024device, bai2024end, asicon}.

CIM architectures implement matrix operations directly within memory arrays, eliminating energy-intensive data movement and offering order of magnitude efficiency improvements over conventional digital systems~\cite{shafiee2016isaac, prime}. However, analog computing elements suffer from inherent device non-idealities that introduce weight perturbations, which can severely degrade neural network inference accuracy and potentially offset the hardware acceleration benefits~\cite{icassp, mao2025hyimc, wang202340nm, pan2022mini}.

The impact of weight perturbations on SSMs is particularly concerning due to their recurrent nature and the importance of precisely tuned parameters for maintaining stable dynamics~\cite{gu2020hippo}. Perturbations in SSM parameters can compound across the sequence length, potentially causing significant performance degradation~\cite{zhoudate}. Despite the growing interest in deploying SSMs on efficient hardware, the robustness of these models to weight perturbations remains largely unexplored~\cite{mamba_quant}.

To bridge this gap, we analyze the vulnerability of SSM components to weight perturbations, identifying that the final block and output projection layers are particularly susceptible. Based on these findings, we propose a lightweight output correction strategy that factorizes the output projection matrix into two components: one implemented on CIM hardware to ensure compatibility with existing hardware architectures, and the other offloaded to digital hardware for precise calibration.
In summary, this paper makes the following key contributions:
\begin{itemize}
    \item We systematically investigate the robustness of SSMs under noisy CIM conditions, revealing that the final block and output projection layers are more vulnerable to perturbations compared to other components.
    \item We propose a novel strategy that factorizes the last output projection matrix into two components: \( \mathbf{U}\mathbf{\Sigma} \) implemented on CIM hardware to ensure compatibility with existing hardware architectures, and \( \mathbf{V^\top} \) offloaded to digital hardware for correction.
    \item Tests on Mamba models show our approach cuts perplexity by up to 99.57\% under noise and boosts PIQA accuracy by up to 96.67\% compared to baselines.
\end{itemize}

\section{Preliminaries}
\label{sec:preliminaries}
SSMs represent a class of sequence modeling architectures that parameterize sequence transformations using linear time-invariant (LTI) systems. An SSM transforms an input sequence $\mathbf{x}(t) \in \mathbb{R}$ into an output sequence $\mathbf{y}(t) \in \mathbb{R}$ through a hidden state $\mathbf{h}(t) \in \mathbb{R}^N$ according to the following continuous-time system:
\begin{align}
\frac{d\mathbf{h}(t)}{dt} &= \mathbf{A}\mathbf{h}(t) + \mathbf{B}\mathbf{x}(t) \\
\mathbf{y}(t) &= \mathbf{C}\mathbf{h}(t) + \mathbf{D}\mathbf{x}(t)
\end{align}
where $\mathbf{A} \in \mathbb{R}^{N \times N}$, $\mathbf{B} \in \mathbb{R}^{N \times 1}$, $\mathbf{C} \in \mathbb{R}^{1 \times N}$, and $\mathbf{D} \in \mathbb{R}$ are learnable parameters. For discrete inputs with time step $\Delta$, this system is converted to its discrete-time equivalent:
\begin{align}
\mathbf{h}_t &= \bar{\mathbf{A}}\mathbf{h}_{t-1} + \bar{\mathbf{B}}\mathbf{x}_t \\
\mathbf{y}_t &= \mathbf{C}\mathbf{h}_t + \mathbf{D}\mathbf{x}_t
\end{align}
where $\bar{\mathbf{A}} = e^{\mathbf{A}\Delta}$ and $\bar{\mathbf{B}} = (\mathbf{A}^{-1}(e^{\mathbf{A}\Delta} - \mathbf{I}))\mathbf{B}$. This discretization allows SSMs to process sequence data efficiently.

Recent SSM variants have improved upon the basic formulation. The S4 model \cite{s4} introduced structured parameterizations of $\mathbf{A}$ using the HiPPO matrix \cite{gu2020hippo}, enabling efficient modeling of long-range dependencies. S4 computes the convolution between input $\mathbf{x}$ and a structured SSM kernel $\mathbf{K}$ using fast Fourier transforms:
\begin{align}
\mathbf{y} = \mathbf{x} * \mathbf{K}, \quad \mathbf{K} = (\mathbf{C}\bar{\mathbf{B}}, \mathbf{C}\bar{\mathbf{A}}\bar{\mathbf{B}}, \mathbf{C}\bar{\mathbf{A}}^2\bar{\mathbf{B}}, \ldots)
\end{align}

Mamba \cite{gu2023mamba} introduced selective state space models, which make the SSM parameters input-dependent:
\begin{align}
\mathbf{A}(\mathbf{x}), \mathbf{B}(\mathbf{x}), \mathbf{C}(\mathbf{x})
\end{align}

This input-dependent parameterization allows the model to selectively process different parts of the input sequence with varying dynamics. In practice, Mamba computes:
\begin{align}
\bar{\mathbf{B}}, \bar{\mathbf{C}}, \Delta &= \text{Projections}(\mathbf{x}) \\
\bar{\mathbf{A}} &= \exp(\Delta \cdot \text{diag}(\mathbf{A})) \\
\mathbf{h}_t &= \bar{\mathbf{A}} \odot \mathbf{h}_{t-1} + \bar{\mathbf{B}} \odot \mathbf{x}_t \\
\mathbf{y}_t &= \bar{\mathbf{C}} \odot \mathbf{h}_t
\end{align}
where $\odot$ represents element-wise multiplication, and the time step $\Delta$ is computed from the input, enabling selective forgetting and retention of information based on the input content.

\section{Methodology}
\label{sec:methodology}
\subsection{Vulnerability of SSMs under Weight Perturbation}
\begin{figure}[!t]
\centering
\includegraphics[width=1.0\linewidth]{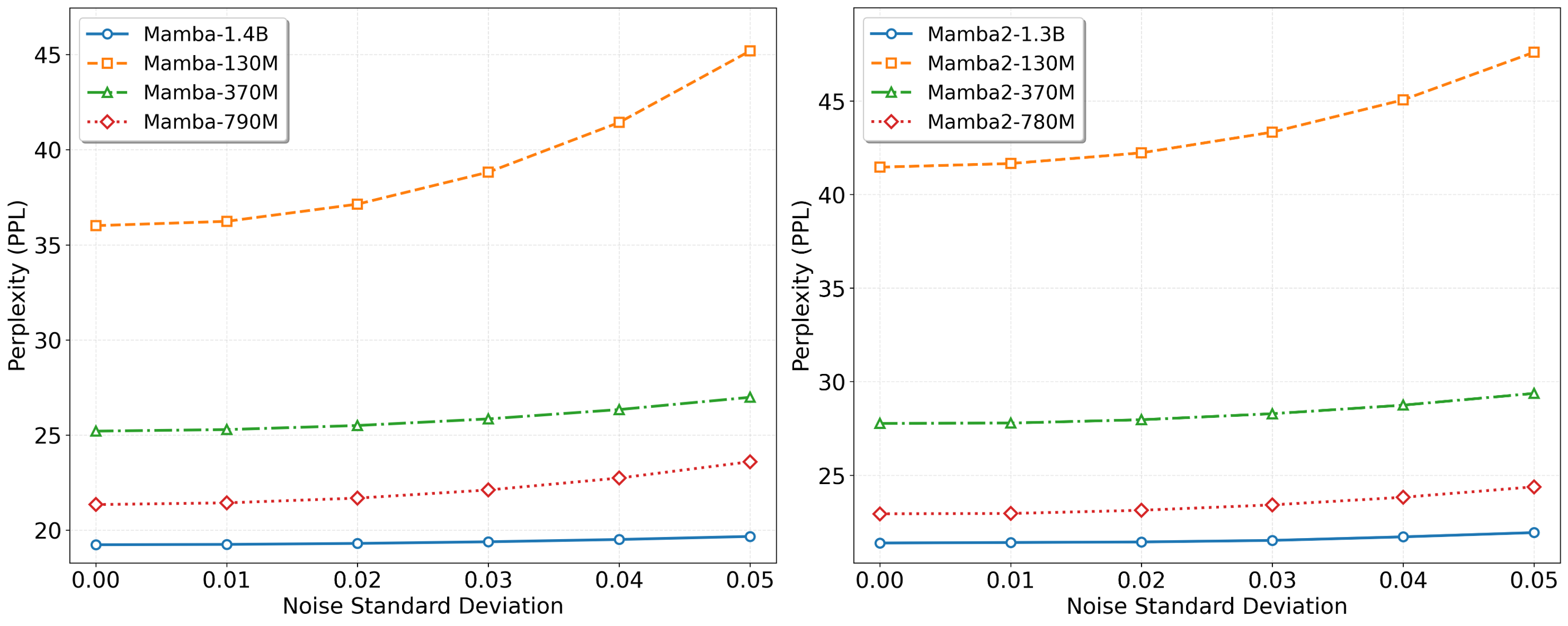}
\caption{Perplexity versus noise standard deviation for Mamba (left) and Mamba2 (right) models of varying sizes. Both model families show increased perplexity with higher noise levels, with smaller models (130M, orange) being most sensitive to noise injection.}
\label{fig:mamba_model_noise}
\end{figure}

Fig.~\ref{fig:mamba_model_noise} illustrates how weight perturbation affects the performance of Mamba models of varying sizes. As noise standard deviation increases from 0 to 0.05, we observe a clear relationship between model size and robustness. The smallest model (Mamba-130M) exhibits the highest baseline perplexity and shows the most dramatic degradation under noise, with perplexity increasing from approximately 36 to 45. In contrast, larger models demonstrate significantly better performance and greater resilience to noise perturbations. A similar trend can be observed in the Mamba-2 series in the right panel. This pattern suggests that parameter count serves as a buffer against noise interference, with larger models maintaining more stable internal representations despite increasing noise levels.

\begin{figure}[!t]
\centering
\includegraphics[width=1.0\linewidth]{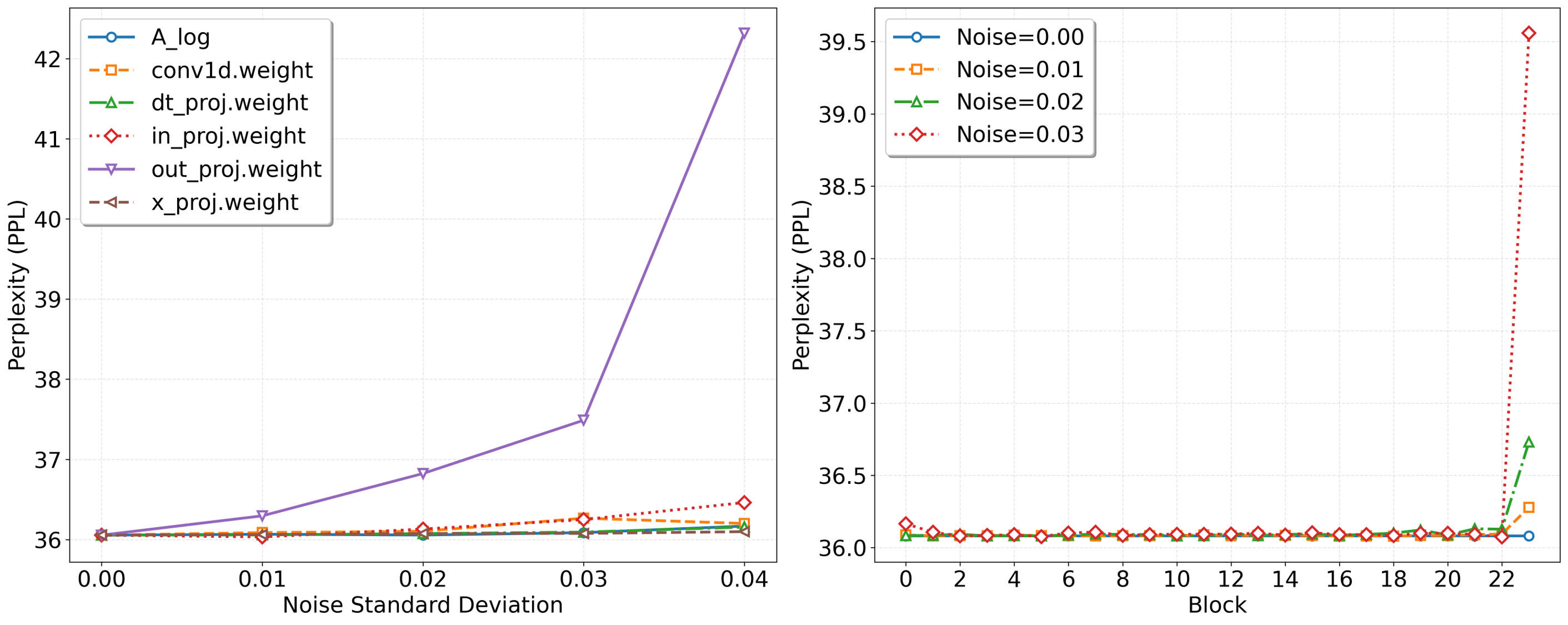}
\caption{(Left) Perplexity versus noise standard deviation for different layer types. (Right) Per-block sensitivity analysis across the 24-layer Mamba-130M model, demonstrating that last blocks exhibit higher sensitivity to noise perturbations.}
\label{fig:mamba_noise_sensitivity}
\end{figure}

Fig.~\ref{fig:mamba_noise_sensitivity} illustrates the robustness of different components within the Mamba architecture when subjected to noise perturbations. The left panel reveals that while the matrix A in SSM mechanism, convolution layer, dt projection, x projection and input projection layer maintain relatively stable perplexity under increasing noise levels, the output projection layer exhibits significant vulnerability, with perplexity rising dramatically at noise standard deviations of 0.03 and 0.04. The right panel demonstrates how noise affects different blocks of the model, showing consistent performance across most blocks but a sharp performance degradation at the final blocks. This block-wise analysis confirms that later stages of the model's computation are particularly susceptible to noise interference.

\subsection{Hybrid Projection Decomposition}
Based on our vulnerability analysis, we identified that the output projection layer is particularly susceptible to weight perturbations in CIM architectures. To address this issue, we propose Hybrid Projection Decomposition (HPD), a novel approach that strategically decomposes the weight matrix to mitigate the effects of hardware non-idealities while maintaining computational efficiency.
\begin{figure}[!t]
\centering
\includegraphics[width=1.0\linewidth]{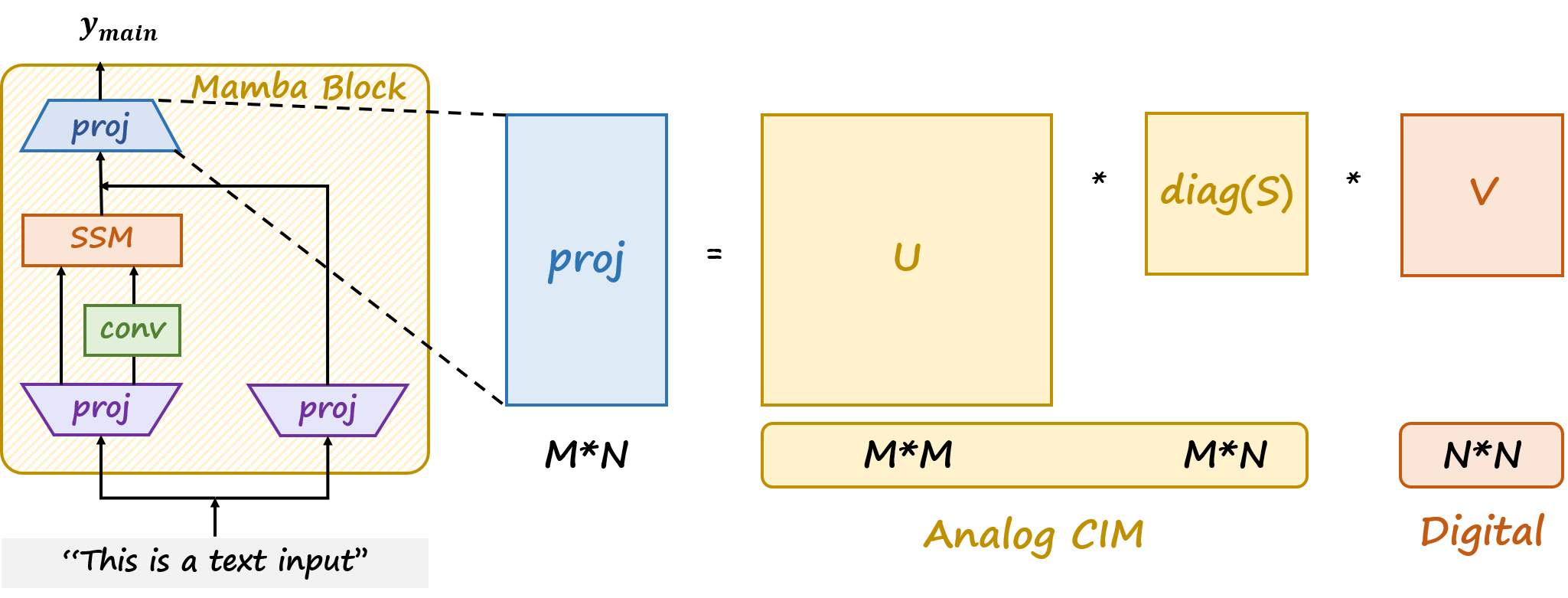}
\caption{An illustration of SSM architectures, combining recurrent and convolutional computations via their linear nature. SSMs enable recurrent inference and parallel training, leveraging RNN and Transformer strengths.}
\label{fig:framework}
\end{figure}

In SSMs, the output projection layer transforms the state representation into the final output space through a linear transformation:
\begin{equation}
\mathbf{y} = \mathbf{W}_{\text{out}}\mathbf{h} + \mathbf{b}
\end{equation}
where $\mathbf{W}_{\text{out}} \in \mathbb{R}^{d_{\text{model}} \times d_{\text{vocab}}}$ is the weight matrix, $\mathbf{h} \in \mathbb{R}^{d_{\text{model}}}$ is the hidden state, and $\mathbf{b} \in \mathbb{R}^{d_{\text{vocab}}}$ is the bias term. 

Our approach leverages Singular Value Decomposition (SVD) to factorize the weight matrix $\mathbf{W}_{\text{out}}$ as:
\begin{equation}
\mathbf{W}_{\text{out}} = \mathbf{U}\mathbf{\Sigma}\mathbf{V}^{\top}
\end{equation}
where $\mathbf{U} \in \mathbb{R}^{d_{\text{model}} \times r}$ contains the left singular vectors, $\mathbf{\Sigma} \in \mathbb{R}^{r \times r}$ is a diagonal matrix with singular values, $\mathbf{V} \in \mathbb{R}^{d_{\text{vocab}} \times r}$ contains the right singular vectors, and $r$ is the rank of the decomposition. 

The key innovation of HPD lies in the strategic distribution of computation between CIM and digital hardware:
\begin{equation}
\mathbf{y} = (\mathbf{V}^{\top})^{\top} [(\mathbf{U}\mathbf{\Sigma})\mathbf{h}] + \mathbf{b}
\end{equation}

We implement this as a two-stage process:
\begin{equation}
\mathbf{z} = (\mathbf{U}\mathbf{\Sigma})\mathbf{h}
\end{equation}
\begin{equation}
\mathbf{y} = \mathbf{V}\mathbf{z} + \mathbf{b}
\end{equation}

The first computation $(\mathbf{U}\mathbf{\Sigma})\mathbf{h}$ is performed on CIM hardware, where we precompute $\mathbf{W}_{\text{CIM}} = \mathbf{U}\mathbf{\Sigma}$ to maintain full compatibility with existing CIM array architectures. This is possible because the product $\mathbf{U}\mathbf{\Sigma}$ preserves the same dimensionality and matrix multiplication structure as the original weight matrix, requiring no modifications to the underlying CIM hardware design or control circuitry. The second stage ($\mathbf{V}\mathbf{z}$) is executed on digital hardware, which provides higher precision and immunity to analog noise.

\section{Experiments}
\label{sec:experiments}
\subsection{Experiment Setup}
All models are implemented using PyTorch 2.2.0 and tested on an NVIDIA L20 GPU. To evaluate performance, we compare our model to its vanilla counterparts under various noise conditions, including Gaussian noise and lognormal noise, due to the lack of relevant research in this area.

\subsection{Experiment Result}
The results in Tables~\ref{tab:gaussian_result} and~\ref{tab:log_result} demonstrate that our method achieves consistent improvements over the vanilla Mamba models under both Gaussian and lognormal noise on the Wikitext dataset. The primary metric is perplexity (PPL), and the robustness ratio is defined as:
\begin{equation}
\text{Ratio} = 1 - \frac{\text{PPL}_{\text{noise, ours}} - \text{PPL}_{\text{no noise, original}}}{\text{PPL}_{\text{noise, original}} - \text{PPL}_{\text{no noise, original}}}
\end{equation}
where \(\text{PPL}_{\text{no noise, original}}\) represents the PPL of the original model under no noise conditions, \(\text{PPL}_{\text{noise, original}}\) is the PPL of the original model with noise, and \(\text{PPL}_{\text{noise, ours}}\) is the PPL of our method under noise. This ratio quantifies how much our method reduces the noise-induced performance degradation relative to the original model.
\begin{table}[!t]
\renewcommand{\arraystretch}{1.3}
\caption{Perplexity (PPL) Comparison of Mamba and Mamba2 Models Across Different Gaussian Noise Levels.}
\centering
\resizebox{\columnwidth}{!}{
\begin{tabular}{llccc}
\toprule
\multirow{2}{*}{Model}      & \multirow{2}{*}{Size} & \multicolumn{3}{c}{PPL under Gaussian Noise} \\ \cmidrule(lr){3-5}
                            &                         & $\sigma = 0.01$      & $\sigma = 0.03$      & $\sigma = 0.05$   \\ \hline
\multirow{4}{*}{\textbf{Mamba}}      & 130M      & 33.64 (+\textcolor{orange}{84.93\%})         & 34.35 (+\textcolor{orange}{88.12\%})           & 35.78 (+\textcolor{orange}{90.34\%})\\
                            & 370M      & 24.58 (+\textcolor{orange}{57.13\%})         & 24.89 (+\textcolor{orange}{12.60\%})           & 25.60 (+\textcolor{orange}{7.37\%})\\
                            & 790M      & 21.90 (+\textcolor{orange}{38.29\%})         & 22.25 (+\textcolor{orange}{57.06\%})           & 22.95 (+\textcolor{orange}{63.71\%})\\
                            & 1.4B      & 19.74(+\textcolor{orange}{36.38\%})         & 19.83 (+\textcolor{orange}{11.32\%})           & 20.04 (+\textcolor{orange}{5.56\%})\\ \cmidrule(lr){1-5}
\multirow{4}{*}{\textbf{Mamba2}}      & 130M      & 37.21 (+\textcolor{cyan}{99.57\%})         & 37.70 (+\textcolor{cyan}{35.14\%})           & 39.04 (+\textcolor{cyan}{25.69\%})\\
                            & 370M      & 26.27 (+\textcolor{cyan}{51.81\%})         & 26.70 (+\textcolor{cyan}{32.44\%})           & 27.57 (+\textcolor{cyan}{25.10\%})\\
                            & 780M      & 22.43 (+\textcolor{cyan}{85.51\%})         & 22.84 (+\textcolor{cyan}{23.17\%})           & 23.69 (+\textcolor{cyan}{22.85\%})\\
                            & 1.3B      & 21.19 (+\textcolor{cyan}{91.53\%})         & 21.19 (+\textcolor{cyan}{97.18\%})           & 21.45 (+\textcolor{cyan}{78.49\%})\\
                            \bottomrule
\end{tabular}
}
\label{tab:gaussian_result}
\end{table}
\begin{table}[!t]
\renewcommand{\arraystretch}{1.3}
\caption{Perplexity (PPL) Comparison of Mamba and Mamba2 Models Across Different Lognormal Noise Levels.}
\centering
\resizebox{\columnwidth}{!}{
\begin{tabular}{llccc}
\toprule
\multirow{2}{*}{Model}      & \multirow{2}{*}{Size} & \multicolumn{3}{c}{PPL under lognormal Noise} \\ \cmidrule(lr){3-5}
                            &                         & $\sigma = 0.01$      & $\sigma = 0.03$      & $\sigma = 0.05$   \\ \hline
\multirow{4}{*}{\textbf{Mamba}}      & 130M      & 33.65 (+\textcolor{orange}{85.64\%})         & 34.37 (+\textcolor{orange}{88.30\%})           & 35.78 (+\textcolor{orange}{90.61\%})\\
                            & 370M      & 24.60 (+\textcolor{orange}{58.38\%})         & 24.90 (+\textcolor{orange}{13.22\%})           & 25.60 (+\textcolor{orange}{7.95\%})\\
                            & 790M      & 21.91 (+\textcolor{orange}{37.36\%})         & 22.27 (+\textcolor{orange}{56.47\%})           & 22.97 (+\textcolor{orange}{62.78\%})\\
                            & 1.4B      & 19.74 (+\textcolor{orange}{35.58\%})         & 19.83 (+\textcolor{orange}{11.16\%})           & 20.05 (+\textcolor{orange}{5.56\%})\\ \cmidrule(lr){1-5}
\multirow{4}{*}{\textbf{Mamba2}}      & 130M      & 37.21 (+\textcolor{cyan}{90.50\%})         & 37.72 (+\textcolor{cyan}{33.92\%})           & 39.11 (+\textcolor{cyan}{24.34\%})\\
                            & 370M      & 26.33 (+\textcolor{cyan}{87.73\%})         & 26.95 (+\textcolor{cyan}{50.41\%})           & 28.08 (+\textcolor{cyan}{22.24\%})\\
                            & 780M      & 22.43 (+\textcolor{cyan}{85.49\%})         & 22.84 (+\textcolor{cyan}{25.00\%})           & 23.68 (+\textcolor{cyan}{38.83\%})\\
                            & 1.3B      & 21.19 (+\textcolor{cyan}{91.65\%})         & 21.18 (+\textcolor{cyan}{91.39\%})           & 21.49 (+\textcolor{cyan}{85.28\%})\\
                            \bottomrule
\end{tabular}
}
\label{tab:log_result}
\end{table}

Under Gaussian noise, as shown in Table~\ref{tab:gaussian_result}, our method consistently improves robustness across both Mamba and Mamba2 models, as evidenced by lower perplexity (PPL) increases compared to their no-noise conditions. For smaller models (\textit{e.g.}, 130M), our method demonstrates significant robustness improvement, achieving PPL increases of 99.57\% and 84.93\% at $\sigma = 0.01$ for Mamba2 and Mamba models, respectively, compared to substantially higher increases in baseline models. As the model size increases (\textit{e.g.}, 1.3B), our method maintains consistent effectiveness for Mamba2, achieving a PPL increase of 78.49\% compared to 90.34\% for the Mamba model at $\sigma = 0.05$. These results highlight the consistent ability of our method to reduce noise-induced degradation across different model architectures and sizes, with the most pronounced benefits observed in smaller models where inherent noise resilience is typically limited. 

For lognormal noise, as shown in Table~\ref{tab:log_result}, our method similarly retains its generalization ability across different statistical noise distributions. For smaller models (\textit{e.g.}, 130M), our method achieves PPL increases of 90.50\% and 85.64\% for Mamba2 and Mamba models, respectively, at $\sigma = 0.01$, demonstrating robust performance even under the heavy-tailed characteristics of lognormal perturbations. For larger models (\textit{e.g.}, 1.3B), PPL increases of 85.28\% and 90.61\% are observed for Mamba2 and Mamba models, respectively, at $\sigma = 0.05$, indicating that the method's effectiveness scales appropriately with model capacity. The consistent improvements across both noise distributions provide strong evidence for the fundamental robustness mechanisms introduced by our approach, rather than overfitting to specific noise characteristics.

\begin{table}[!t]
\renewcommand{\arraystretch}{1.3}
\caption{Evaluation of Accuracy for Mamba Models Under Gaussian Noise.}
\centering
\resizebox{\columnwidth}{!}{
\begin{tabular}{llccc}
\toprule
\multirow{2}{*}{Model}      & \multirow{2}{*}{Size} & \multicolumn{3}{c}{Accuracy (\%)} \\ \cmidrule(lr){3-5}
                            &                         & ARC-e      & PIQA & LAMBADA   \\ \midrule
\multirow{4}{*}{\textbf{Mamba}}      & 130M      & 47.73 (+\textcolor{orange}{77.78\%})         & 64.24 (+\textcolor{orange}{96.67\%})           & 42.39 (+\textcolor{orange}{57.22\%})\\
                            & 370M      & 54.55 (+\textcolor{orange}{37.50\%})         & 69.22 (+\textcolor{orange}{66.67\%})           & 53.65 (+\textcolor{orange}{27.50\%})\\
                            & 790M      & 60.98 (+\textcolor{orange}{59.09\%})         & 72.36 (+\textcolor{orange}{50.00\%})           & 61.23 (+\textcolor{orange}{66.98\%})\\
                            & 1.4B      & 65.26 (+\textcolor{orange}{9.52\%})          & 75.06 (+\textcolor{orange}{33.33\%})           & 65.08 (+\textcolor{orange}{36.51\%})\\
\bottomrule
\end{tabular}
}
\label{tab:common}
\end{table}

To verify the generalization ability of our method, we evaluate Mamba models of varying sizes under Gaussian noise perturbations on commonsense reasoning datasets (Table~\ref{tab:common}). Our approach demonstrates significant improvements in robustness across all model sizes, with particularly substantial gains observed in the ARC-e benchmark: 77.78\% improvement for the 130M model, 37.50\% for the 370M model, and 59.09\% for the 790M model. The PIQA benchmark shows consistent improvements ranging from 33.33\% to 96.67\%, demonstrating enhanced physical commonsense reasoning capabilities under noise conditions. Similarly, the LAMBADA benchmark exhibits improvements from 36.51\% to 66.98\%, confirming the method's effectiveness across diverse reasoning tasks and linguistic contexts.

\section{Conclusion}
\label{sec:conclusion}
This work presents the first systematic study of weight perturbations in SSMs on analog CIM hardware, pinpointing the output projection layers in the final block as highly noise-sensitive. Our HPD method splits the output projection matrix into CIM-compatible and digitally calibrated components for better precision. Tests on Mamba models show HPD cuts perplexity by up to 99.57\% under noise and boosts PIQA accuracy by up to 96.67\% for commonsense reasoning, compared to baselines.
\section*{Acknowledgement}
\label{sec:acknowledgement}
This research was partially conducted by ACCESS – AI Chip Center for Emerging Smart Systems, supported by the InnoHK initiative of the Innovation and Technology Commission of the Hong Kong Special Administrative Region Government, and partially supported by the Theme-based Research Scheme (TRS) project T45-701/22-R, the General Research Fund (GRF) Project 17203224 of the Research Grants Council (RGC), Hong Kong SAR, and the National Natural Science Foundation of China Project 62404187.
\bibliographystyle{ieeetr}
\bibliography{arxiv}
\end{document}